# Decomposing the "strange attractor like" seismic electric precursor into simpler components.

Thanassoulas[1], C., Klentos[2], V., Verveniotis, G.[3], Zymaris, N.[4]

1. Retired from the Institute for Geology and Mineral Exploration (IGME), Geophysical Department, Athens, Greece.
   e-mail: thandin@otenet.gr - URL: www.earthquakeprediction.gr

2. Athens Water Supply & Sewerage Company (EYDAP),
   e-mail: klenvas@mycosmos.gr - URL: www.earthquakeprediction.gr

3. Ass. Director, Physics Teacher at 2nd Senior High School of Pyrgos, Greece.
   e-mail: gver36@otenet.gr - URL: www.earthquakeprediction.gr

4. Retired, Electronic Engineer.

**Abstract.**

An attempt is made in this work to decompose the "strange attractor like" seismic electric precursor into more simple and elementary components. The basic data files of the orthogonal **(NS, EW)** components of the Earth's electric field used for the compilation of the corresponding phase maps are decomposed by a joint non-linear inversion scheme into two basic oscillating electric fields. The first one, called "signal", is attributed to a single current source while the second, called "noise", is attributed to the mix-up of some regional and randomly located current sources. The comparison of the phase maps compiled from the raw data files to the ones compiled by the "signal" and "noise" data shows that the newly compiled "strange attractor like" phase maps preserve their predictive property while their appearance resembles simpler geometrical shapes (pure hyperbolas and ellipses). Moreover, it is postulated that its generating mechanism is the stress waves applied in the regional area by the combined interaction of the Sun, Moon and Earth (tidal waves). The latter mechanism, when the seismogenic area is under critical stress-charge conditions, triggers new current sources at places where the stress has reached locally critical stress levels and hence the "strange attractor like" seismic precursor is generated.

## 1. Introduction.

An inspection of the appearance of the "strange attractor like" seismic electric precursor (Thanassoulas 2007, Thanassoulas et al. 2008, 2008a, 2009, 2009a) reveals that the specific attractor takes very peculiar shapes, either as variable or distorted ellipses or as more complicated hyperbolas. The latter, specifically, may be considerably distorted assimilating a random mix of hyperbolas. In this work, we will make an attempt to resolve the seismic "strange attractor like" electric precursor into some simpler and more pure components. Since the "strange attractor like" precursor is calculated from the oscillating component of the Earth's electric field, it is reasonable to focus our efforts on the details of this specific oscillating electric field. The latter is the result of the band-pass filtering operation applied on the raw data registered by the monitoring sites used to record the variations of the Earth's electric field.

Generally, the method which is mostly used by the scientific community, for any data series filtering, is the traditional **Fast Fourier Transform (FFT).** In a brief description of this methodology, the original data are, initially, converted into its frequency spectrum. A filter (of any required kind), is used to retain the frequency band of interest, rejecting the entire unwanted spectrum. Finally, from the retained frequency spectrum, the filtered data, in time domain, are reconstructed. A detailed analysis of the topic was presented by Bath (1974) and Kulhanek (1976). The **FFT** filtering process is presented in the following figure **(1)**. Subfigure **(a)** represents a time function from which a specific monochromatic component must be extracted. Therefore, figure **(a)** is transformed into its frequency spectrum that is represented by subfigure **(b)**. In **(b)**, the monochromatic component of interest **(F1)** is shown too. Next, a very sharp band-pass filter is applied on the frequency content **(b)** of **(a)**, which is represented in **(c)** and finally the wanted monochromatic output is presented in **(d)**.

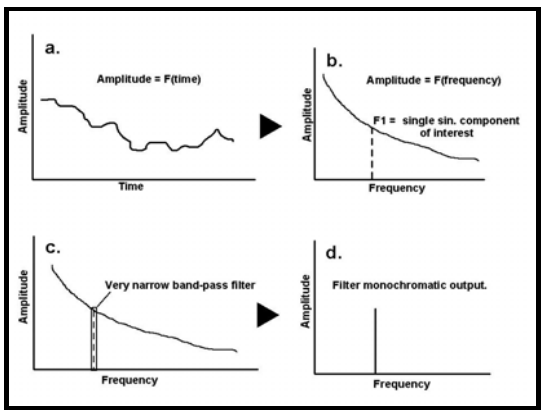

Fig. 1. Schematic presentation in the frequency domain of the band-pass filtering process. **a:** raw time function, **b:** frequency content of **(a)** and monochromatic component of interest **F1, c:** application of a sharp band-pass filter, **d:** filter monochromatic output.

Although the filtering process seems simple, considering a monochromatic output there is a peculiar complication. If in the raw data more than one components of the same frequency of interest but of different phases exist, then, all these different in phase components will pass through the band-pass filter. The latter is demonstrated in figure **(2)**.

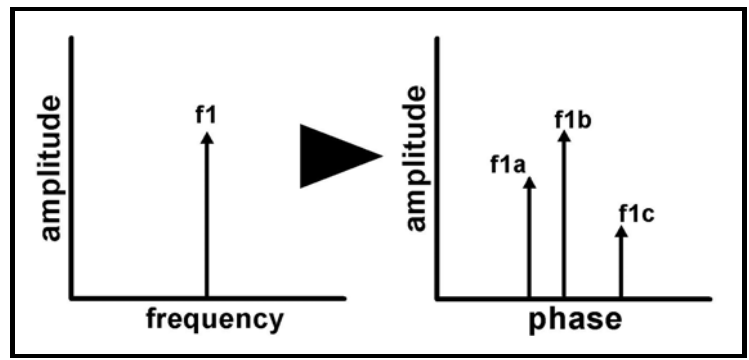

Fig. 2. Monochromatic signal output of the band-pass filtering process represented into phase domain. Left: filter output. Right: different in phase monochromatic components of the same frequency are shown.

In order to facilitate the understanding of this problem a synthetic example is presented in sketch form. In figure **(3a)** a monochromatic oscillating field is present while in figure **(3b)** the same field has been phase-shifted at a certain amount. Both fields **(3a)** and **(3b)** have been combined and are presented in **(3c)** while a direct comparison is made of **(c)** to **(a)** at **(d)**.

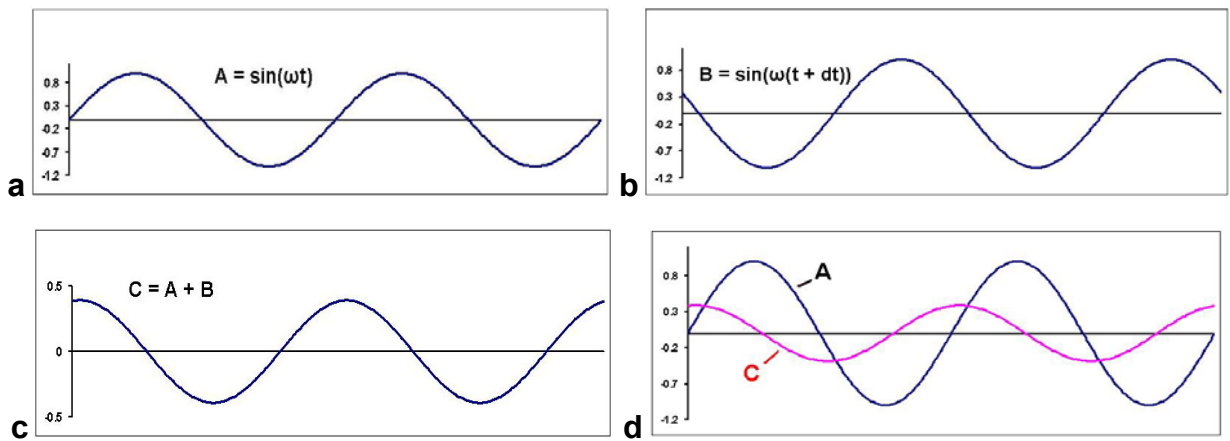

Fig. 3. Comparison of a monochromatic single phase electric field to the result of its infection by a same monochromatic electric field but of different phase, **a:** monochromatic oscillating field, **b:** field of **(a)** has been phase-shifted at a certain amount, **c:** combination of **(a)** and **(b)**, **d:** comparison of **(a)** to **(c)**.

The real importance of the above analysis will be revealed at the case when two orthogonal components of the Earth's electric field will be used for the determination of the azimuthal direction of its intensity vector.

Firstly, the case of a "clear" signal, that is a single in-phase signal, will be considered. In this case the calculated azimuthal direction is theoretically represented by a straight line (Thanassoulas, 2007). In the following figure **(4)** a monochromatic single in-phase two orthogonal components electric field is presented at left, while its electric field intensity vector azimuthal direction is presented at right.

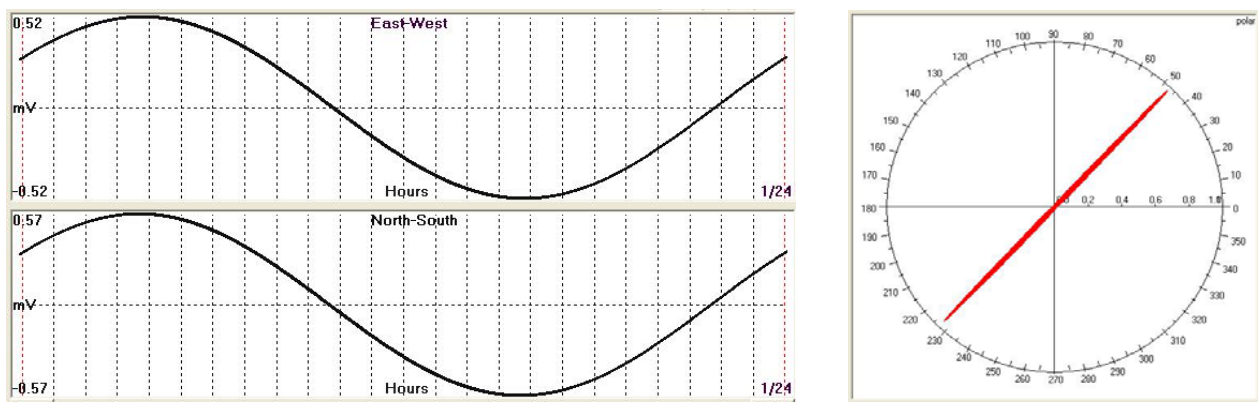

Fig. 4. Monochromatic single phase two orthogonal components electric field (left) and its intensity vector azimuthal direction (right).

The fact that the used data of figure **(4)** come from processing of real raw data, that is they do not correspond to a pure sine wave, has as an effect, in to the corresponding azimuthal direction, to cause a slight deviation of it in respect to a pure straight line.

Secondly, the very same presentation will be made for the case of an electric field that contains some different out of phase same monochromatic components. The data of this case, presented in figure **(5, left),** show clearly the phase-shift between the two orthogonal components while its azimuthal direction presented in figure **(5, right)** is no more a straight line, on the contrary it consists of rather two "fat" ellipses.

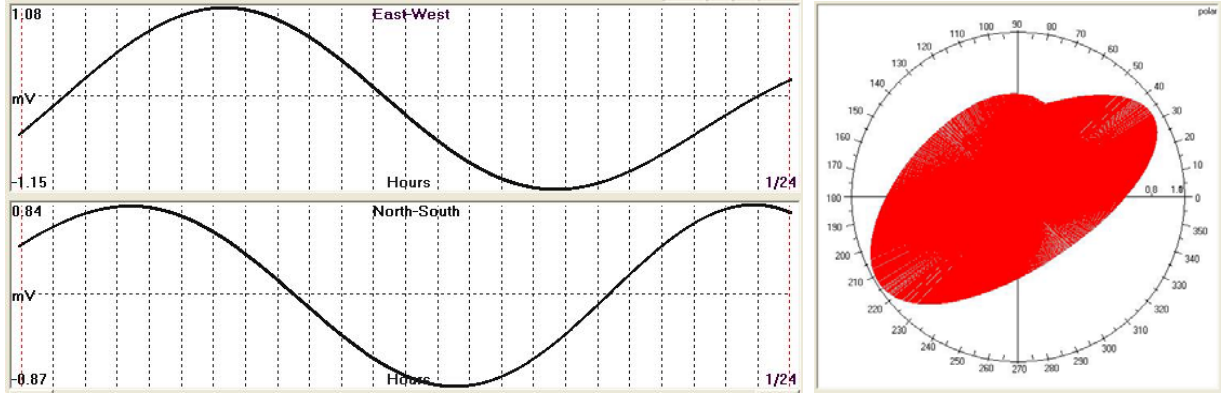

Fig. 5. Monochromatic out of phase two orthogonal components electric field (left) and its intensity vector azimuthal direction (right).

Similar data, as the ones presented in figure **(5),** were used in order to calculate the "strange attractor like" seismic electric precursor (Thanassoulas 2007, Thanassoulas et al. 2008, 2008a, 2009, 2009a).

The questions that immediately arise are: is it possible to separate the different in phase components of the monochromatic electric field into distinct ones from each other? How the "strange attractor like" seismic precursor is affected from these different in phase components of the monochromatic electric field? Is it possible to decompose it into more simple components?

This work tries to answer these questions.

## 2. Theoretical analysis.

It is considered that the "strange attractor like" seismic electric precursor is generated, somehow, by the interaction of two electric fields. The first one is the generated, by any valid physical mechanism, electric signal by the focal area of the pending earthquake. The second electric signal is generally the induced electric noise which can be generated by any kind of none / or seismic process. This mix-up of electric signals holds for both orthogonal components **(E-W, N-S)** of any monitoring site that records the Earth's electric field. Consequently, the data which are used for the calculation of the "strange attractor like" seismic electric precursor, for each monitoring site, contain the following components:

    **a. E-W dipole:**

        **Preseismic electric signal at time ti:** $V_{S1}\sin(\omega t_i + t_1)$     (1)

        **Noisy electric signal at time ti**   : $V_{N1}\sin(\omega t_i + t_2)$     (2)

    And the recorded **(E-W)i** of the total electric field is:

    **(E-W)i = $V_{S1}\sin(\omega t_i + t_1) + V_{N1}\sin(\omega t_i + t_2)$**     (3)

    Where: $V_{S1}$   is the peak value of the **EW** preseismic signal
       : $V_{N1}$   is the peak value of the **EW** noisy signal
       : $t_1$     is the phase of $V_{S1}$
       : $t_2$     is the phase of $V_{N1}$

Similarly, the electric field that is registered by the **N-S** dipole consists of the following components:

    **b. N-S dipole:**

        **Preseismic electric signal at time ti:** $V_{S2}\sin(\omega t_i + t_1)$     (4)

        **Noisy electric signal at time ti**   : $V_{N2}\sin(\omega t_i + t_3)$     (5)

    And the recorded **(N-S)i** of the total electric field is:

    **(N-S)i = $V_{S1}\sin(\omega t_i + t_1) + V_{N1}\sin(\omega t_i + t_2)$**     (6)

    Where: $V_{S2}$   is the peak value of the **NS** preseismic signal

: $V_{N2}$   is the peak value of the **NS** noisy signal
: $t_1$      is the phase of $V_{S2}$
: $t_3$      is the phase of $V_{N2}$

At this point it is reminded that the preseismic signal recorded in both **E-W** and **N-S** dipoles have the same phase **t1** due to the fact that these signals are generated by the same physical mechanism in the focal area of the pending earthquake (Thanassoulas 1991, 2007). The mathematical model that prescribes the actual pass-band filtered raw data recorded by the two dipoles is a time **(ti)** function of: $V_{S1}, V_{N1}, V_{S2}, V_{N2}, t_1, t_2,$ and **t3** parameters.

The raw pass-band filtered data will be analyzed into "signal" and "noise" accordingly. The two pair of signals (signal and noise) will be used separately to calculate the corresponding "strange attractor like" seismic electric precursor. For the purpose of the wanted analysis we used the Marquardt (1963) and Figueroa (1980) method which is a well known and widely used one in geophysical problems, for least squares estimation of non-linear parameters.

Typically, the method is presented in the following figure **(6)**. The inputs of the Marquardt procedure are: a) a data series of observations and b) the model's starting parameters. It is assumed that the model accounts for the generation of similar data as the input observations. The Marquardt procedure optimizes the starting input parameters of the model so that the value of RMS error between the observed data series and the generated ones by the optimized model is at minimum.

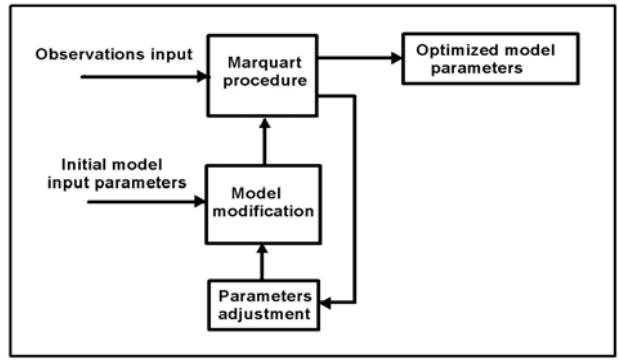

Fig. 6. Typical Marquardt procedure sketch presentation for the least squares estimation of non-linear parameters.

A classical Marquardt algorithm was modified so that it could accommodate (joint inversion) as input simultaneously **E-W, N-S** band-pass filtered data and the model parameters earlier presented. During a test-run of the program a real data set **(fig. 7)** was analyzed into signal **(fig. 8, left)** and noise **(fig. 8, right)** components and then the original signal was reconstructed from both these components **(fig. 9)**. Finally a comparison is made between the original input data and the calculated ones from the optimized model.

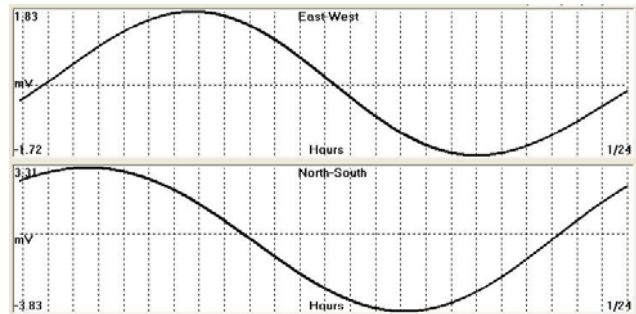

Fig. 7. Observed **E-W / N – S** band-pass **(T=24h)** filtered raw data.

The data set of figure **(7)** was decomposed **(fig. 8)** into signal (left) and to noise signal (right).

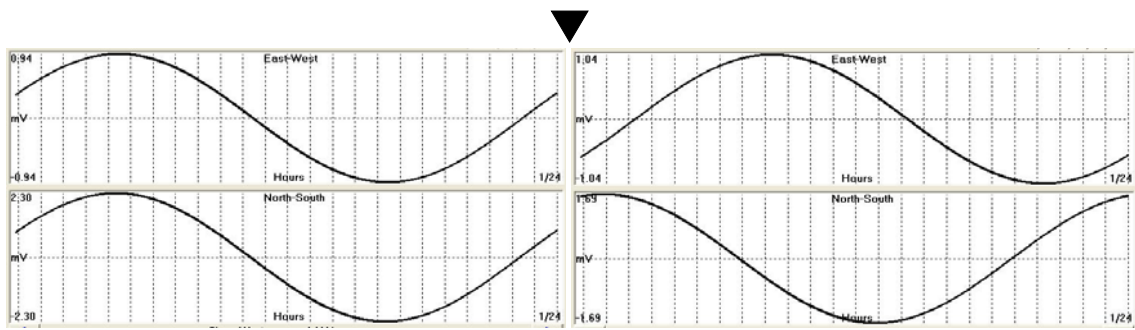

Fig. 8. Left: observed signal data, Right: observed noise data

At the final step of testing this procedure a comparison is made between the input data and the calculated ones (synthesis of signal and noise) by the optimized model.

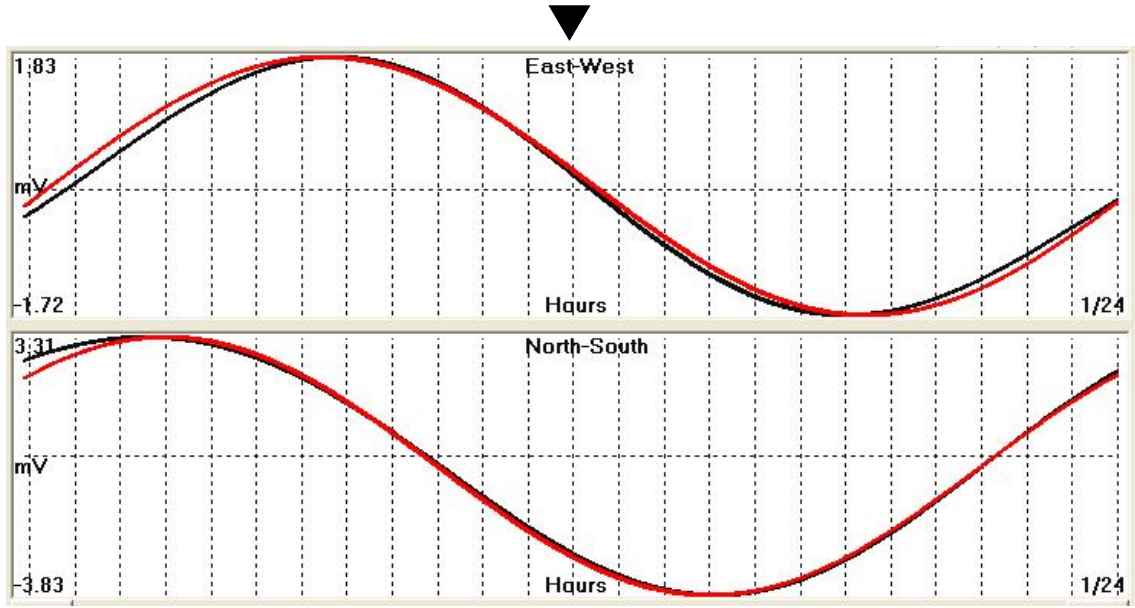

Fig. 9. Comparison of observed (black) to synthesized (red) data.

The already presented procedure will be applied on real data obtained from **PYR** and **HIO** monitoring sites (Thanassoulas, 2007). Then the obtained signals and the corresponding noise will be used to calculate the "strange attractor like" seismic electric precursor for two distinct dates. The first one is the 18[th] February of 2008, when no seismic precursor is present **(fig. 10, 11, 12).** A complicated drawing of mixed hyperbolas characterizes the "strange attractor like" seismic precursor. The second is the 8[th] of February 2008, that is a few days before the large Methoni EQ (**Ms=6.7R**) of the 14[th] February, 2008. A set of double ellipses characterizes the "strange attractor like" seismic precursor **(fig. 20).**

Following are presented the raw oscillating data, its corresponding azimuthal direction of the Earth's electric field intensity vector, the corresponding "strange attractor" map, the decomposed components (preseismic signal – noise) and the relevant azimuthal direction diagrams along with their corresponding "strange attractor" maps.

## 3. "Strange attractor like" phase map (hyperbolas) of 18[th] February, 2008.

This example refers to a date when no "strange attractor like" seismic precursor was observed. The phase map is characterized by the presence of mixed random hyperbolas. Therefore it represents the typical behavior of the Earth's electric field under no severe preseismic excitation at all.

**Raw data (PYR)**

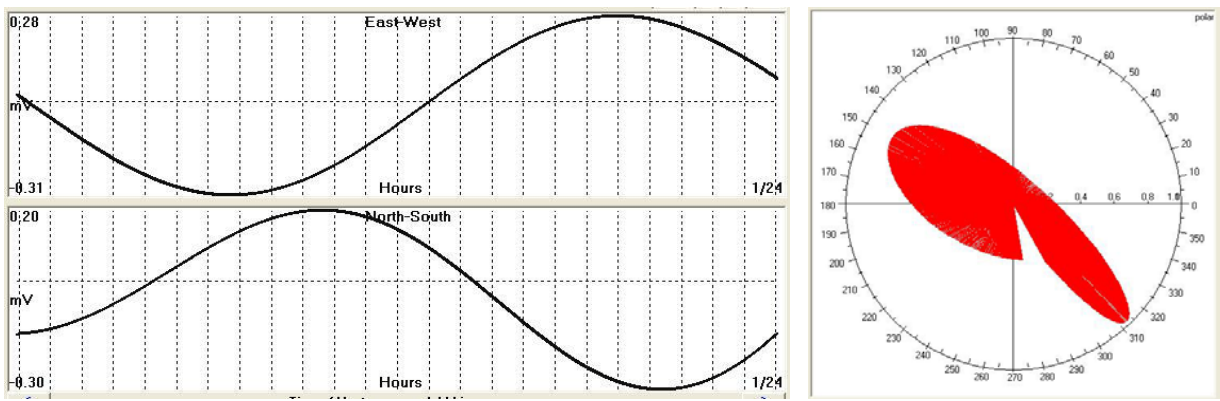

Fig.10. Left: raw data, Right: azimuthal direction **(PYR 080218).**

**Raw data (HIO)**

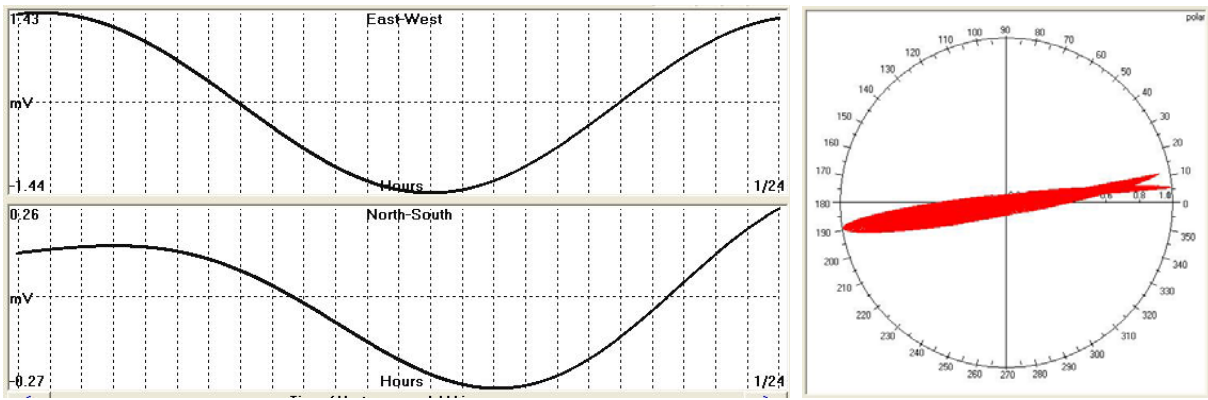

Fig.11. Left: raw data, Right: azimuthal direction **(HIO 080218)**.

Both polar diagrams of the Earth's electric field intensity vector calculated for **PYR** and **HIO** monitoring sites show a large deviation from the straight line. Using these data sets, from both monitoring sites, the following **(fig. 12)** "strange attractor like" phase map was compiled

**Corresponding "strange attractor like" phase map.**

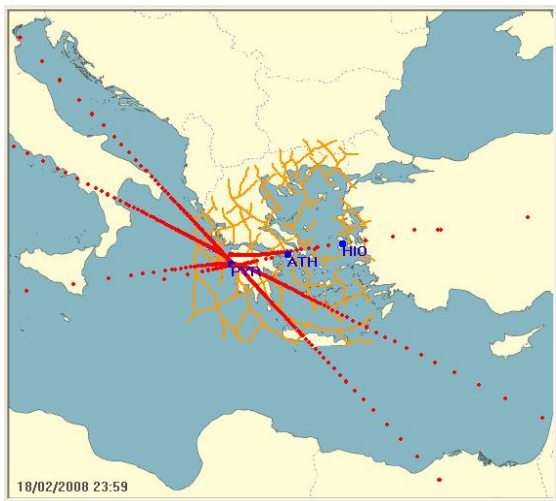

Fig. 12. "Strange attractor like" phase map calculated from data of figures **10, 11 (PYR – HIO, 080218).**

The compiled "strange attractor like" phase map consists of a small number of mixed random hyperbolas.
Next, the data presented in figures **(10, 11)** were processed (in a joint inversion scheme) and analyzed into in-phase "signal" and out of phase "noise" components as follows:

**Processed data, signal (PYR).**

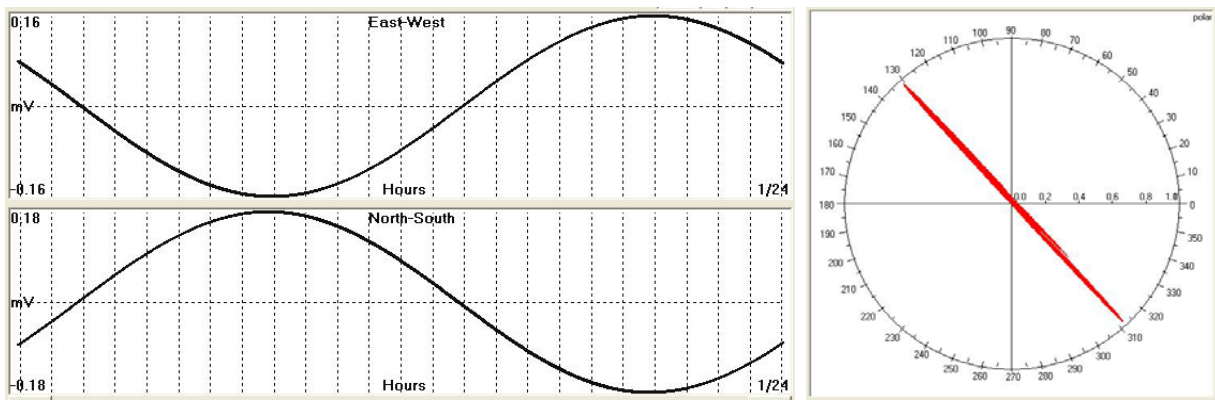

Fig. 13. Left: signal, Right: azimuthal direction **(PYR 080218)**.

**Processed data, noise (PYR).**

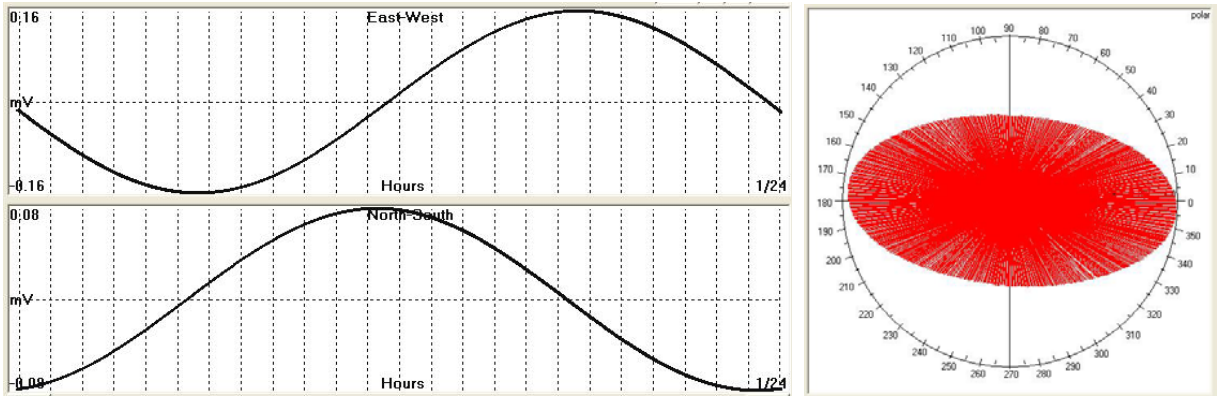

Fig. 14. Left: noise, Right: azimuthal direction **(PYR 080218).**

**Processed data, signal (HIO).**

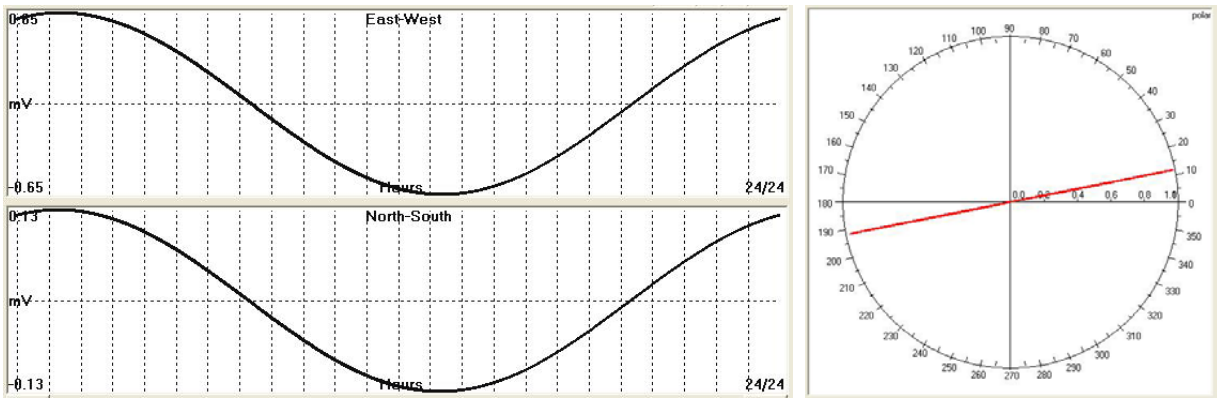

Fig. 15. Left: signal, Right: azimuthal direction **(HIO 080218).**

**Processed data, noise (HIO).**

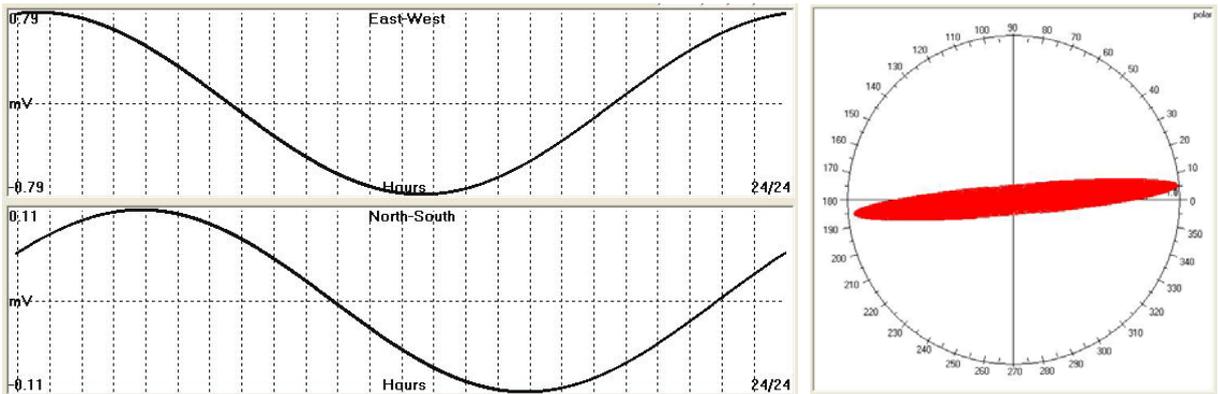

Fig. 16. Left: noise, Right: azimuthal direction **(HIO 080218).**

It is made clear from figures **(13, 15)** that the "signal" in-phase components generate a straight line polar diagram, while the "noise" out of phase components generate rather fat regular ellipses.

The data of figures **(13, 14, 15, and 16)** were used to compile the corresponding "strange attractor like" phase maps.

**Corresponding "strange attractor like" phase maps.**

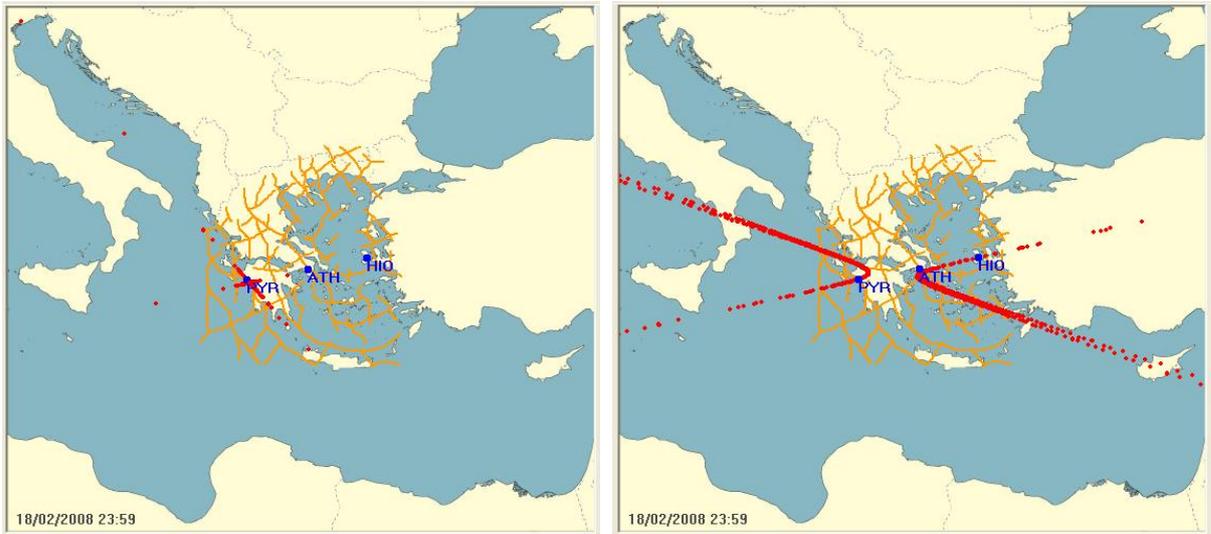

Fig. 17. Left: signal, Right: noise "strange attractor like" phase maps calculated from the data of figures **(13. 14, 15, 16) (PYR – HIO, 080218).**

The "strange attractor like" phase map **(fig. 17, left)** compiled from the "signal" in phase components presents the expected intersection of both azimuthal straight lines calculated at each monitoring site. The observed scattering of some intersections is probably due to the fact that a certain small level of noise still exists in the "signal" data. The "strange attractor like" phase map **(fig. 17, right)** compiled from the "noise" out of phase components presents a pair of well behaved hyperbolas.

## 4. "Strange attractor like" phase map (ellipses) of 8$^{th}$ February, 2008.

This example refers to a date when a "strange attractor like" seismic electric precursor was observed. The raw phase map is characterized by the presence of ellipses. Therefore, it represents the typical behavior of the Earth's electric field under worth to be considered preseismic electric excitation.

**Raw data (PYR)**

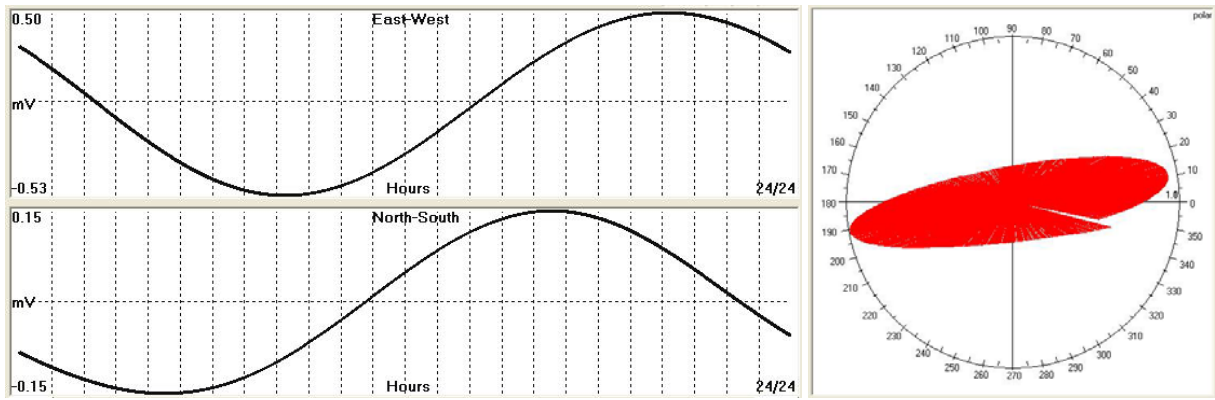

Fig. 18. Left: raw data, Right: azimuthal direction **(PYR 080218).**

**Raw data (HIO)**

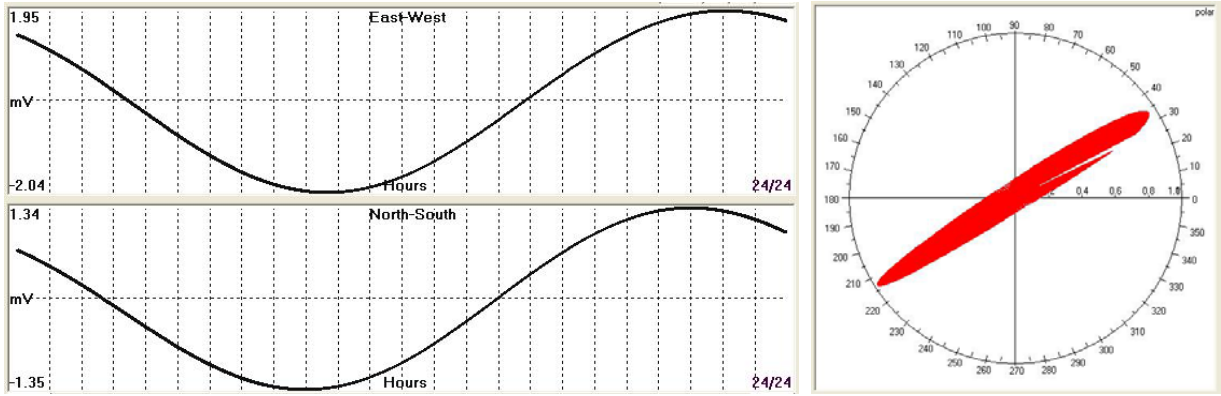

Fig. 19 Left: raw data, Right: azimuthal direction **(HIO 080218).**

As in the previous case, the raw data generate polar diagrams presented as rather fat ellipses. However, in this case the "strange attractor like" phase map represents a pair of ellipses shown in figure **(20).**

**Corresponding "strange attractor like" phase map.**

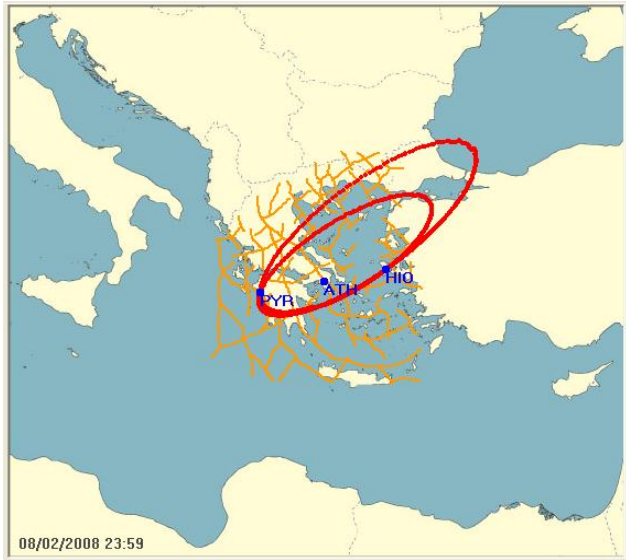

Fig. 20. "Strange attractor like" phase map calculated from data of figures **18, 19 (PYR – HIO, 080208).**

In this case the "strange attractor like" phase map suggests increased electrical activity leading towards a severe seismic event (14[th] February, 2008, Ms = 6.7R). The analysis of the data of figures **(18, 19)** gave the following results:

**Processed data, signal (PYR).**

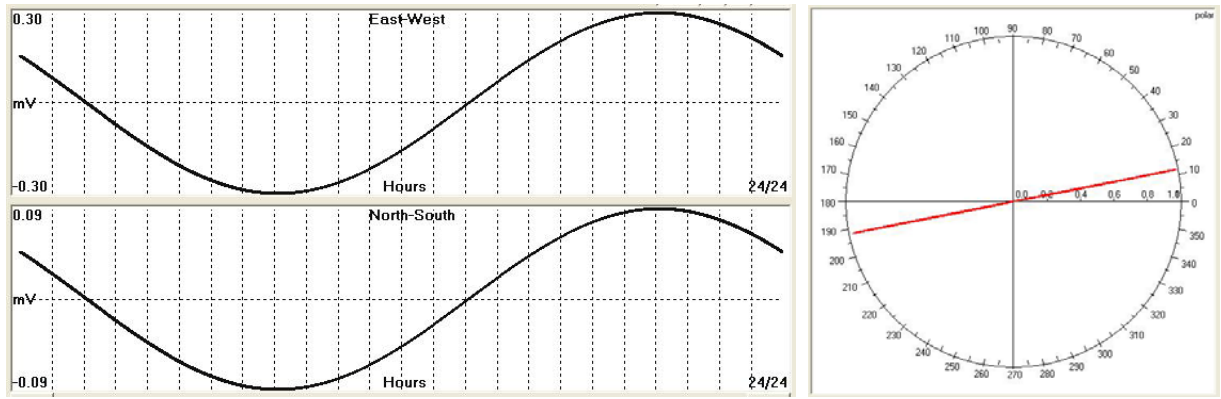

Fig. 21. Left: signal, Right: azimuthal direction **(PYR 080208)**.

**Processed data, signal (HIO).**

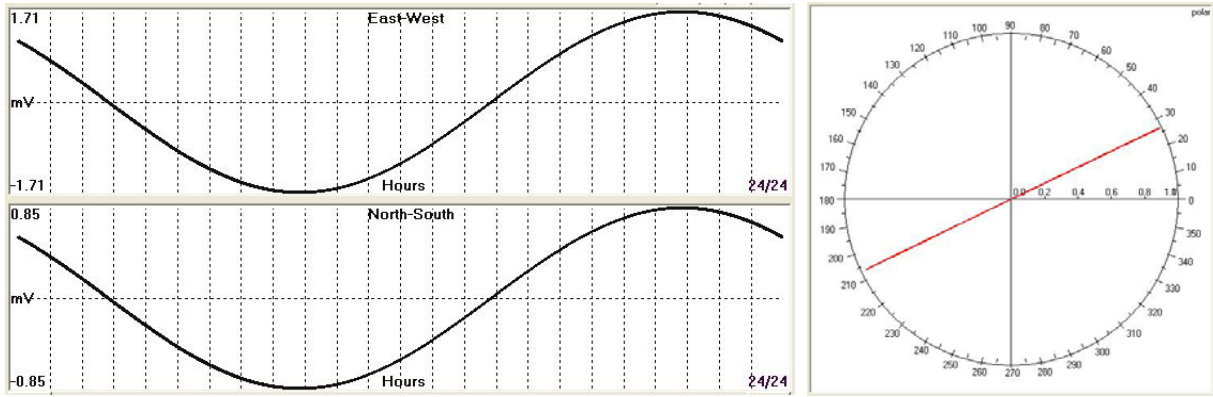

Fig. 22. Left: signal, Right: azimuthal direction **(HIO 080218)**.

**Processed data, noise (PYR).**

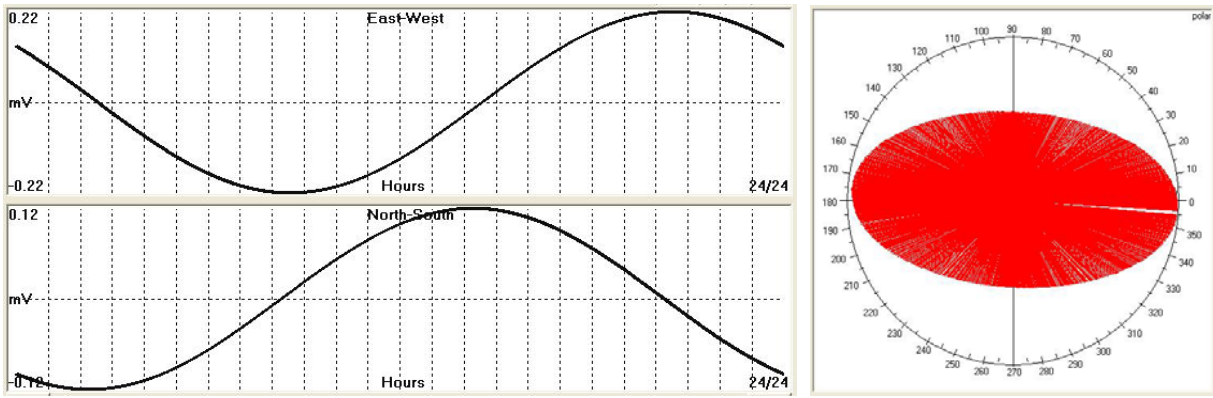

Fig. 23. Left: noise, Right: azimuthal direction **(PYR 080208).**

**Processed data, noise (HIO).**

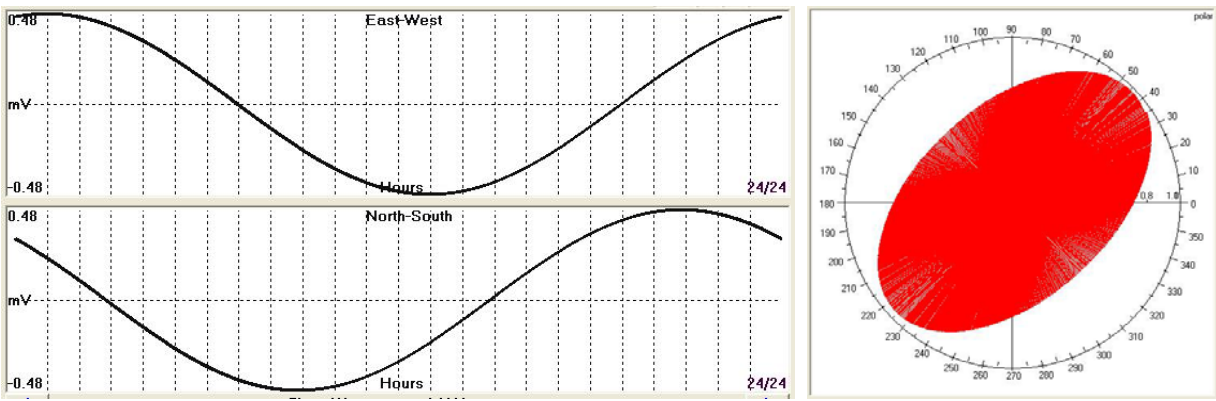

Fig.24. . Left: noise, Right: azimuthal direction **(HIO 080218).**

**Corresponding "strange attractor like" phase maps.**

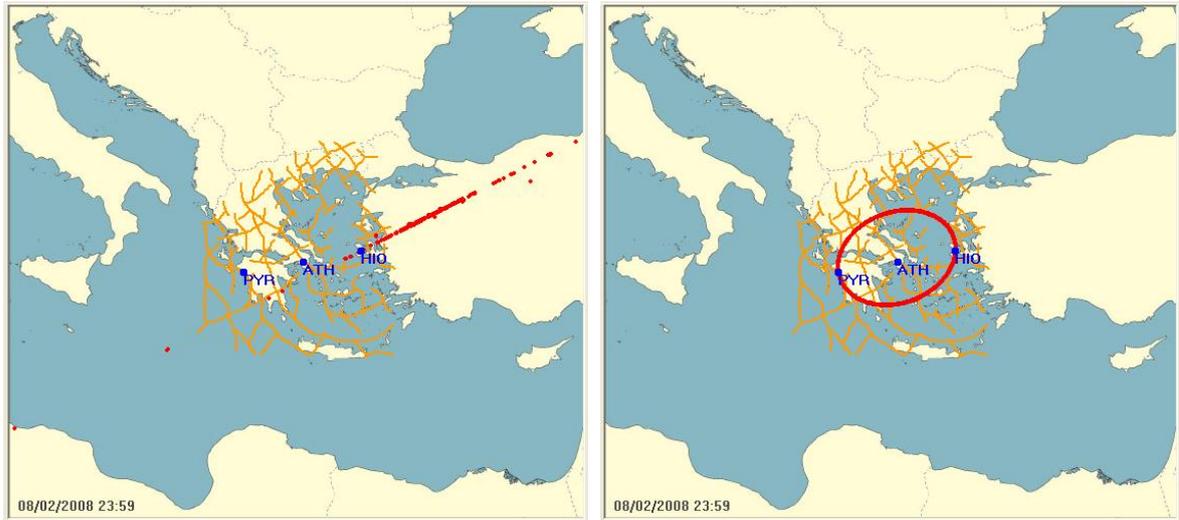

Fig. 25. Left: signal, Right: noise "strange attractor like" phase maps calculated from data
of fig**. 21, 22, 23, 24 (PYR – HIO, 080208).**

Figure **(25, left)** represents the typical intersection of the azimuthal directions of the Earth's electric field intensity vector. The scattering of the intersections around a central point and along a straight line is probably due to small level noise still present in the "signal" data. Figure **(25, right)** represents a nice ellipse.

## 5. Discussions – conclusions.

The concept of the complex Earth's oscillating electric field was firstly introduced by Thanassoulas (2007). Furthermore, it was shown that the polar diagram generated by the two orthogonal components of the Earth's electric field registered at a single monitoring site exhibits directional properties which correspond to the various seismically active regions surrounding the monitoring site (Thanassoulas et al. 2008b) In this work, the oscillating Earth's electric field, of a single monitoring site, was simulated by two oscillating components. The first one consists of an "in-phase" pair of orthogonal components called "signal" while the second consists of an "out of phase" one called "noise". An example of this analysis is presented in figures (**7, 8, 9)**. The small discrepancy observed between the observed and simulated data probably is due to the very simple adopted model (1 in-phase, 1 out of phase component and probably to some small level noise present in the processed data). The use of a more complicated (more in phase and out of phase components) model could rectify this problem. The calculated polar diagrams show the typical straight line for the corresponding "signal" that it is assumed as it is generated by a single electric current source, and wide ellipses or hyperbolas for the "out of phase, noise" orthogonal components, generated from more than one current sources.

The data of both cases were analyzed into "signal" and "noise" components. The latter were used to reconstruct the corresponding "strange attractor like" phase maps presented, for comparison purposes, in the following figures **(26, 27).** Firstly, the one of 18$^{th}$ February, 2008 is presented.

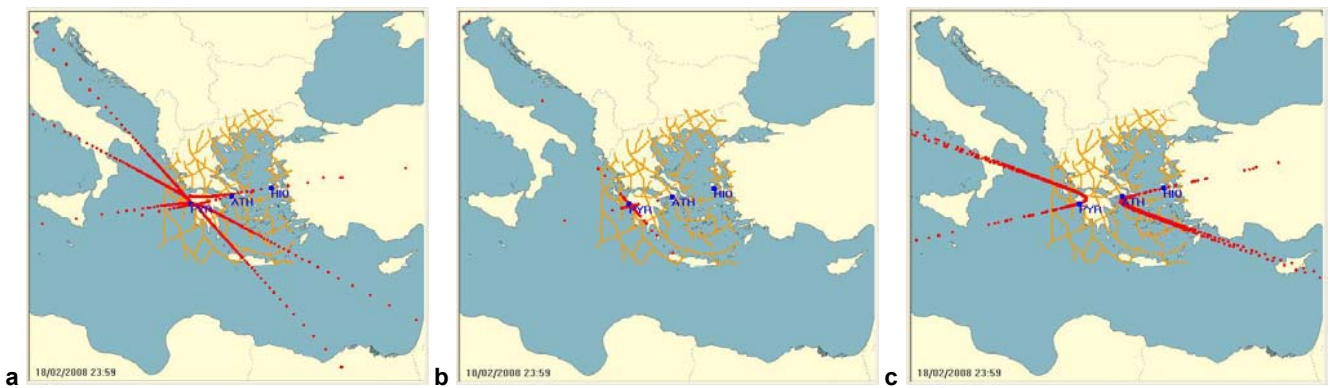

Fig. 26. "Strange attractor like" phase maps compiled from: **a** = raw band-pass **(T=24h)** filter output, **b** = "signal" component, **c** = "noise" component **(PYR – HIO, 080218).**.

In figure **(26a)** is presented the "strange attractor like" phase map that corresponds to the raw oscillating data from both **(PYR – HIO)** monitoring sites. This figure shows the mixed hyperbolas mentioned earlier. In figure **(26b)** the "strange attractor like" seismic electric precursor is represented by the intersection of two straight lines that correspond to the azimuthal directions calculated from the signal of both monitoring sites. This figure fits the theoretical model of a single current source. In figure **(26c)** the "strange attractor like" seismic electric precursor is represented by well-behaved "clean" hyperbolas, calculated from the noise, thus suggesting that no seismic excitation is present.

Secondly, the case of 8$^{th}$ February, 2008 is presented in figure **(27).** This case is quite different from the previous one.

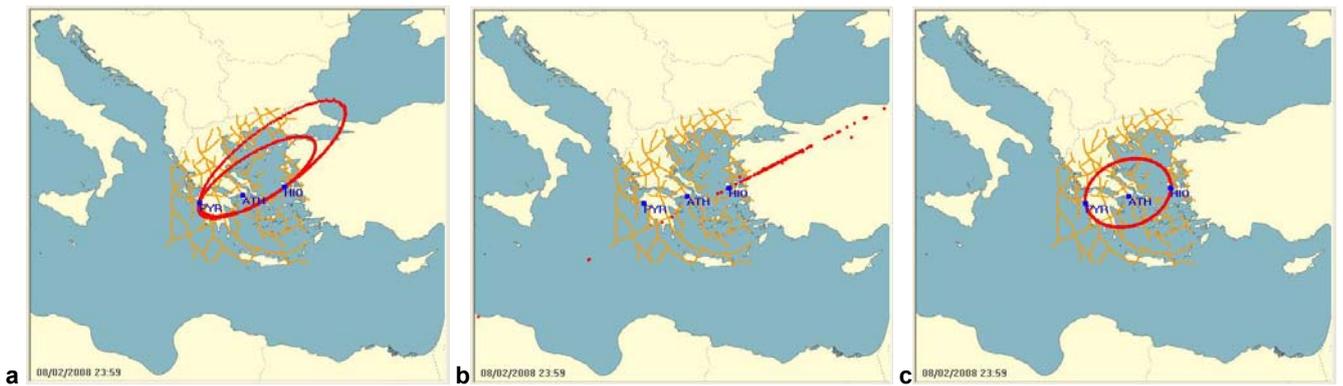

Fig. 27. "Strange attractor like" phase map compiled from: **a** = raw band-pass **(T=24h)** filter output, **b** = signal component, **c** = noise component **(PYR – HIO, 080208)**.

In figure **(27a)** the "strange attractor like" seismic precursor suggests that a severe EQ is under preparation (the one of 14th February, 2008, Ms = 6.7R). In figure **(27b)** the "strange attractor like" seismic precursor is represented by the intersection of two straight lines that correspond to the azimuthal directions calculated from the signal of both monitoring sites. This figure fits the theoretical model of a single current source. The linear spread of the intersections around a central point is due to the fact that a small amount of noise still exists in the "signal" in phase calculated data. Finally, in figure **(27c)** the "strange attractor like" seismic precursor is represented by well-behaved "clean" ellipse calculated from the noise, thus suggesting that some severe seismic excitation will show up soon.

It is made clear that the "strange attractor like" seismic electric precursor shape can be qualitatively improved by its analysis into "signal" and "noise" components. Furthermore, its capability to indicate a non excited (hyperbolas) or an excited (ellipse) seismically regional area, is not changed at all after this analysis.

An advantage of this analysis is the fact that in both cases of hyperbolas or ellipses the signal "strange attractor like" seismic electric precursors can be used for further mathematical analysis due to the fact that are pure geometrical shapes. The case of the "signal" phase map (intersection of straight lines like azimuthal directions) could be used for further refinement of the epicentral area of a pending large EQ. Especially the case of the ellipse is very interesting since it is produced by "out of phase" orthogonal components. Therefore, it is suggested that its generation mechanism is not a single current source. Rather it consists of the mix-up of various same frequency and different in phase signals, generated from different current sources (focal areas). This is corroborated by the multidirectional properties of the Earth's electric field found at the recordings of a single monitoring site (Thanassoulas et al. 2008b). The latter could lead to the identification of the location of these current sources before the main seismic event takes place.

Finally, a word must be said about the triggering mechanism of all these different current sources required to produce the "strange attractor like" seismic electric precursor. The proposed model works as follows: In order to activate this precursor it is required that more than one current sources must have simultaneously been triggered. Therefore, a wide regional stress load increase is required. The only one mechanism that is capable to produce such an effect is a "stress wave" that temporarily stress loads the regional seismogenic area. The latter are the well-known tidal waves generated by the Sun – Earth – Moon gravitational interaction. Consequently, the "strange attractor like" seismic electric precursors must be closely correlated to the corresponding tidal waves (Thanassoulas et al. 2008, 2008a, 2009, 2009a).

This type of analysis could be successively reapplied on the "noise" data for obtaining finer details of the original "strange attractor like" seismic electric precursor compiled from the raw data.

## 6. References.